\documentclass[prE,twocolumn,showpacs]{revtex4}
\usepackage[T1]{fontenc}
\usepackage[cp1250]{inputenc}
\usepackage{color}
\usepackage{graphicx}
\usepackage{hyperref}

\begin{document}

\title{Studies of opinion stability for small dynamic networks with opportunistic agents}

\author{Pawel Sobkowicz}
\affiliation{Warsaw, Poland}
\email{pawelsobko@gmail.com}
\homepage{http://countryofblindfolded.blogspot.com}
\thanks{Preprint of an article accepted by \emph{International Journal of Modern Physics C}. Copyright World Scientific Publishing Company.}

\begin{abstract}
There are numerous examples of societies with extremely stable mix of contrasting opinions. We argue that this stability is a result  of an interplay between society network topology adjustment and opinion changing processes. To support this position we present a computer model of opinion formation based on some novel assumptions, designed to bring the model closer to social reality.
In our model, the agents, in addition to changing their opinions due to influence of the rest of society and external propaganda, have the ability to modify their social network, forming links with agents sharing the same opinions and cutting the links with those they disagree with. To improve the model further we divide the agents into `fanatics' and `opportunists', depending on how easy is to change their opinions. The simulations show significant differences compared to traditional models, where network links are static. In particular, for the dynamical model where inter-agent links are adjustable, the final network structure and opinion distribution is shown to resemble real world observations, such as social structures and persistence of minority groups even when most of the society is against them and the propaganda is strong. 
\end{abstract}


\maketitle

\section{Introduction}

Computer modeling of social behavior  is becoming increasingly popular part of sociophysics, discipline combining statistical methods and physical insights with sociological observations. A recent review article\cite{1} lists  seventeen reasons for using such models, the three topmost being prediction, explanation and guiding data collection. The models, optimally starting with simple, understandable and common sense premises should allow us to understand at least some social phenomena.

One of the directions of such research is a study of dynamics of opinion distribution in societies. 
The literature on the subject is quite rich, with several popular approaches, for example Sznajd model\cite{2}, Deffuant model\cite{3} or Krause-Hegelsmann model\cite{4}. Other approaches have been proposed by Nowak and Latan{\'e}\cite{5} and Kacperski and Hołyst\cite{6}. Most of these and related works focus on the mechanisms allowing the agents to come to a common opinion, either via direct interactions in encounters between pairs of agents or through general influence of society as a whole. The social structure itself may be very different, from traditional one or two-dimensional geometry to complex networks. For a review  of these models see Ref.~\citep{7}, for additional discussion of possible directions see also Ref.~\citep{8}. The main interest of opinion formation studies is in finding conditions which lead to a society reaching unanimous opinion, either via internal interactions or thanks to external influences. For this reason the field is often called `consensus formation'.

On the other hand, the real world examples of societies where full consensus is present are quite rare. 
In fact, where  important issues are at stake, we observe extremely stable situations of opinion splits, ranging from relative parity (for example in political setup of many countries) to configurations where small minority persists despite strong social pressures (for example in terrorist support cases).

 Our goal is to study  opinion formation taking into account some processes that are usually omitted, which, in our opinion play a crucial role in real social situations and increase the the stability of split opinion configurations. We study changes of opinion in conjunction with other processes, such as segregation. Current `standard' approaches to opinion formation modeling focus on changes of individual opinions due to encounters between agents. It is surprising that the tendency of voluntary social separation due to attitude or opinion similarities and differences, well known and recorded even in a folklore (`birds of a feather group together') has been neglected in opinion simulations so far. Instead of the choice of withstanding the arguments of an opponent or surrendering to them, most of us have the choice of severing the relationships with them. This results in relatively closed sub-societies, with little interaction between them, 
especially when the issues under study are of crucial value. A classical example of  social modeling study invoking the separation tendencies is the famous Schelling model of social segregation\cite{9}.

A novel aspect of the computer model of opinion formation presented here is that agents have the ability to cut the social links with those they disagree with and  form a new links with agents sharing the same opinion. This changes the network structure. To take into account the fact that in real societies some links can not be broken (e.g. family or work relationships) we have simulated situations where certain percentage of the links remain \emph{static}, while the rest are \emph{free}, allowing changes in topology of the social network.

 To keep the model simple, we have assumed binary opinion values ($+, -$), which may describe some highly contested social issues. Examples include political preferences and opinions on topics such as abortion or evolution. 
Within the model, the strength of influence between agents decreases with their social separation, reflecting the fact that our opinions are swayed less by remote acquaintances or strangers than by the closest associates. Secondly, the opinion of a given agent may be changed in reaction to perceived cumulative  social opinion of others, corresponding to properly averaged `peer pressure' rather than the individual encounters. Many of the classical models have stressed the importance of such individual contacts in opinion changes of agents, but in our belief the constant background of perceived opinions, resulting from numerous meetings (not necessarily related to the issue in question) and information on opinions held by other society members is more relevant. In a way, this can be described as each agent continuously measuring and responding to the `discomfort' due to difference between its opinion and properly averaged opinions of other agents. In addition to this intrinsic pressure we have added simple parametrized external influence, aimed to describe propagandist efforts.

As a further enhancement of the model, we have divided the agents into two classes: `fanatics' and `opportunists'. The only difference is that it is relatively hard to change the opinion of a fanatic. They correspond to core `believers' in each of the alternative opinions, for example the extreme `pro-' or `anti-abortion' activists or political party members.  The opportunists correspond to the soft support base of a given opinion and exhibit much greater probability of `following the trend', adjusting their own opinion to the perceived majority or to external propaganda. 
We have assumed that the fanatics form a small part of  the society (10\%), parameter which is obviously adjustable, and look for effects extending beyond their number.
One of the questions we have asked was when is it possible that a set of fanatics holding a minority opinion can influence sizable number of supporters. This is rather important in situations of terrorism, where the existence  of the core of active terrorists depends on a wider circle of supporters.

We have based our social model on  scale-free Albert-Barabási (AB) networks.  Such networks reproduce an important effect found in many natural and artificial systems, including social relationships, namely the existence of a large fraction of very highly connected network nodes (for general reviews see, e.g. Refs.~\citep{10,11,12,13}). In the AB networks the dispersion of the number of immediate neighbors (degree distribution) scales according to power law rather than exponential or Gaussian. This type of networks results from interplay of two processes: growth of the network and preferential attachment (the `rich get richer' principle). This proves to be very important factor in opinion formation. Should `fanatic' agents be placed in such `positions of influence', their impact on the general behavior is  more pronounced \cite{14}.

\section{Computer simulation model}
\label{SecSimul}

The main features of the model are as follows:
\begin{itemize}
	\item A set of $N$ interacting agents forms a society. The dynamics of the total system is assumed to take place through discrete time steps. 
		\item Each agent $i$ has, at a given time, his `opinion' $\sigma(i)$ ($+$ or $-$). The initial number of $+$ opinion holders, $N_+^{ini}$ is one of the crucial control parameters determining the results of the simulations.
	\item Some of the links are assumed to be \emph{static}, corresponding to family or workplace connections, which are difficult to break even if there is a difference in opinions of the connected agents. The rest of the links are \emph{free}, and may be cut off when the opinions of the agents are different.  New links may form between agents sharing the same opinion (and are assumed to be free).  
To preserve the general network properties the process of destruction and creation of links is performed in a way that keeps the  number of connections constant.  
	\item Each agent is characterized by its opportunism. This is modeled via parameter $\omega(i)$ ($0\leq \omega(i) \leq 1$). Small values of $\omega(i)$ correspond to agents with low probability of changing their opinion under the influence of others, or external influences -- describing the `fanatic' agents. High values of $\omega(i)$ correspond to `opportunistic' agents who readily change their opinion to join majority or follow external propaganda. The model allows any distribution of $\omega(i)$, but the results presented in this work were obtained in a simple binary model, where the `fanatics' have $\omega(i)=\omega_f$ and `opportunists' have $\omega(i)=\omega_o$, ($\omega_f<<\omega_o$).
	\item  Influence of agents on other agents is governed by their respective opinions  as well as on their social distance. We measure this distance simply via the number of links separating the agents in the network.  
\end{itemize}	

	The simulation program consists of the following major steps:	
\begin{enumerate}
	\item Initial formation of a network, assignment of initial opinion $\sigma(i)$ and opportunism $\omega(i)$ values and 	initial assignment of the  of static and free status to the links of the network.  The average number of links should be realistic, and for results presented here we have chosen, on the average, 6  bidirectional links per agent. 
	\item As the starting distribution of all the above mentioned parameters is random, it has little connection with real societies, with their correlated structures and patterns of opinions. To keep the model closer to reality, we start with a relatively small number of iterations ($<5000$), during which  the agents adjust the network links but are not allowed to change their opinions. This produces society network already separated due to differences in opinions.  Starting simulations with a random distribution (quite typical for many social simulations) has been found to give spurious results. For simulations where all the links are assumed static, this phase simply does not change anything.
	\item The following, much larger number of iterations ($<200000$), includes two steps: possible opinion change for a randomly chosen agent and adjustment of network connections by the same agent via cutting off or creation of a new link.
\end{enumerate}
These two key processes of the simulation are described below.

For a randomly chosen `initializing' agent $i$ we calculate  the network separation for all other agents and separate them into four spheres $S_1, S_2, S_3$ and $S_4$. These are composed of, respectively, the direct neighbors, the second neighbors, third neighbors and the rest of the society. The strength of the influence of an agent $j$ on $i$ depends on the sphere that $j$ belongs to. For each agent $j$ we calculate also a modification factor $f_{\sigma}(i,j)$, which reflects our human tendency to give more value to opinions of those who agree with us and to give less credibility to contrary opinions. Thus $f_{\sigma}(i,j)=2$ if $\sigma(i)=\sigma(j)$ and $f_{\sigma}(i,j)=1$ otherwise.

For each of the spheres $S_K$ we calculate the averaged, normalized influence over agent $i$
\begin{equation}
I_K = \frac{- \sum_{j\in S_K} f_{\sigma}(i,j) \sigma(i) \sigma(j)}{\sum_{j\in S_K} f_{\sigma}(i,j)}
\end{equation}
The denominator acts to normalize $I_K$, so that if all agents $j$ within $S_K$ disagree with $i$, then $I_K$ is at the upper limit of $I_K=1$ and if all agents in $S_K$ agree whit $i$ then $I_K=-1$.

The overall influence of the society on agent $i$ is then calculated as a weighted sum of $I_K$. To reflect the observation that close acquaintances  have much more influence than strangers we have assigned geometrically decreasing weights, for the spheres of increasing distance. The total influence of the society on agent $i$ is 
\begin{equation}
s(i) = \sum_{K=1}^{4} I_K d^{(K-1)} \frac{1-d}{1-d^4},
\end{equation}
where $d<1$ and the last factor serves to provide normalization. In this work we have used $d=0.5$ and $d=0.3$, which correspond, respectively, to 53\% and 70.5\% of the weight of the influence originating from the immediate sphere of neighbors $S_1$.

In addition to social pressure we $s(i)$, we have included a simple model for external influence $h(i) = -h \sigma(i)$, due to media, propaganda or cultural traditions. Here, $h$ is a simple numerical parameter, which, along with the $N_+^{ini}$ is one of the key controls of the simulated system.

The initializing agent $i$ changes its opinion with probability of $p(i)=\left(s(i)+h(i)\right)\omega(i)$. Obviously, when the sum of $s(i)+h(i)$ is lower than zero, there is no chance of opinion change. This corresponds naturally to a situation where enough agents agree with $i$ (making the $s(i)$ negative), and when the external influence has the same sign as $\sigma(i)$. Conversely, $p(i)$ may become positive if enough agents disagree with $i$ (especially agents close to it, from the nearest spheres of influence) or if a strong external propaganda has the opposite sign than $i$'s opinion.

It is worth noting that in this model, we take into account influences of all society members, scaled down in value accordingly to their distance from the initializing agent. The individual contributions of agents belonging to $S_4$ sphere are very small, due to normalization within the sphere and then to the geometric decrease factor for the sphere itself. They form an averaged background to more varied local influences of neighbors from closest spheres.

The second step in simulations is the modification of network connections. 
To maintain the  number of links constant we have alternated cutting and adding links. The first of these processes is very simple. Keeping the previously chosen agent $i$, we pick another, `target' agent $t$ randomly from the set of nearest neighbors that disagree with $i$. The link between $i$ and $t$ is then cut off. Obviously, if all $i$'s neighbors agree with it, then we move to next simulation round.

Creating new links is more complex. The target agent $t$ is not chosen totally at random among the spheres $S_2, S_3$ and $S_4$, but rather with probability  decreasing on social distance between $i$ and $t$. 
This probability for agent $t$ belonging to sphere $S_K$ ($K>1$) around $i$ is given by
\begin{equation}
p(t) = \frac{A}{N_K^{agr}}{\left( \frac{1}{2} \right)}^{K-1}  ,
\end{equation}
where $N_K^{agr}$ is the number of agents in the $S_K$ sphere who agree with $i$ and $A$ is a global normalization factor to ensure that $\sum_t p(t) = 1$.
The $(1/2)^{K-1}$ factor leads to a probability of forming a new ling with agents from further spheres decreasing geometrically.
Again, we chose this form for conceptual simplicity, with the aim of getting a simple equivalent to the observations that we meet more often with our close social neighbors. The newly formed link between $i$ and $t$ is treated as a free one.  
 The resulting network preserves, with only minor deviations, the general degree distribution characteristic for AB networks. However, the process leads very quickly to separation of subnetworks of $+$ and $-$ opinion holders if the free links are dominant. 
The final shape of such separated network resembles very much some distributions found in social studies. 
While the average number of links per agent remains unchanged  and only minor changes in degree distribution are observed, other relevant network properties change radically, for example the shortest path distribution and clustering coefficient. This is especially important when all the links are free, because $+$ and $-$ opinion holders segregate to separate communities. If we calculate the network properties of these communities separately, then in the case of large majority we observe that there is similarity to the original configuration in such characteristics as clustering coefficient. On the other hand, the minority has much larger fraction of disconnected agents, so that there is significant difference in related network characteristics. 

The computer program used for the simulations stores the vital society parameters (such as $N_+$), numbers of links between various types of agents, cumulative number of opinion changes and rewiring events. We also record specific network data at the beginning of the iterations, at the point when the initial phase of network segregation ends and at the end of the simulation. The number of iterations has been chosen in a way to provide a stabilized configuration, by direct monitoring the number of opinion switches and network rewires per 10000 individual steps. It should be noted that thanks to forced nature of rewiring, this part of the process is much faster than the opinions change. For moderate values of $o_f$ we observe also small `cascades' of changes brought by `conversion' of a fanatic.

\section{Analysis of results}

The simulations presented in this paper are run with relatively small artificial societies of 500 agents. This limitation decreases, obviously, the value of our results for general societies. We believe, however, that even small simulation sizes might be found useful to describe at least some social situations, such as local communities or ad-hoc virtual groups found frequently on the Internet.
We have used the average of 6 direct links per agent, and varied the proportion of static ($S$) and free ($F$) links, thus a notation $F=2, S=4$ would denote simulation with, on average, there would be 2 freely re-linkable connections and 4 unchangeable ones per agent. Of course, by the very nature of AB networks,
there would be agents with significantly more  connections.

Our first observation was that for some range of simulation parameters, the results of individual runs were quite unpredictable. When the simulations  stabilized the system often ended up in configurations where the basic property $N_+$  varied greatly. For a wide range of external influence values and the initial split of opinions $N_+^{ini}$,  the final society  could contain 300 $N_+$ agents, or, in another run, it could end up with 490 such agents out of 500. These differences were traced down to particular allocation of positions of fanatic agents, especially when they initially occupied sites of high connectivity.
We have welcomed this unpredictability of the simulations at the level of individual 
runs as a sign that the model has some chances to provide insights into real societies, 
where individual choices and varying influences may have great impact 
on general society situation.

The similarity with reality is largely due to the dynamic nature of the network, 
allowing the structures to adjust to opinions rather than forcing the opinions to adjust 
to the network structure.
One can compare a typical outcome of a simulation that resulted in a mixed opinion 
society with data on blogging behavior  in strongly polarized environment of US 2004 elections, 
which has been published by Adamic and Glance\cite{15}. The observed linking patterns between 
the blogs is almost identical to those obtained in our simulations in the case when all links 
are free for rewiring. It is worth noting that blogging is
a perfect example of social network where such freedom of association and disassociation 
is present, without restrictions imposed by workplace, family of geographical constraints. 
The opposing groups become nearly separated, with very few links between them. 
Even in the case when some static links are present, (for example $S=2, F=4$) the separation of the groups is very strong. 
 
\begin{figure*}
\includegraphics[scale=0.45,angle=-90]{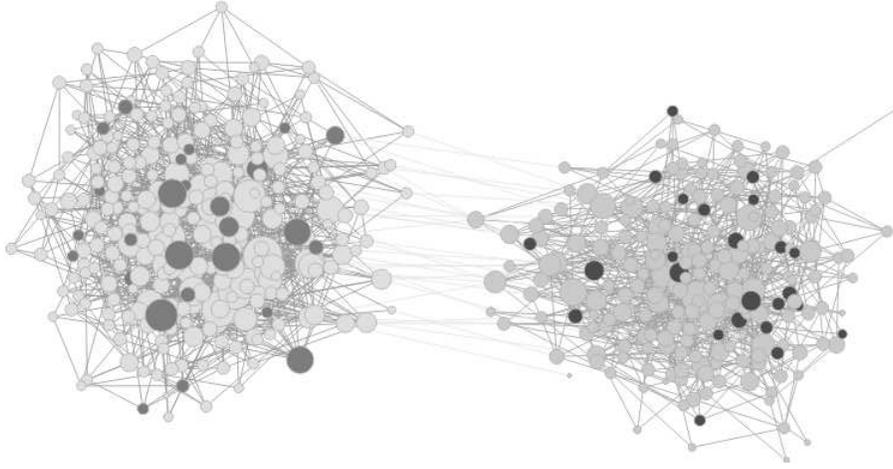} 
\caption{Example of a network in  a model society, with relatively few links between them. We have used automatic program algorithm optimizing agent positioning, which resulted in a clear division between the agents with $+$ and $-$ opinions. Darker circled denote the fanatics, and the size of the circles corresponds to number of the links of an agent. This simulation corresponds almost ideally to observations of political behavior described by  Adamic and Glance.  \label{networkS0}  
}
\end{figure*}

Because the results of individual simulations were quite unpredictable, 
we have concentrated on the statistical properties of the simulation outcomes. 
The basic measured quantity was the final number of $+$ agents, $N_+$.
We have looked for conditions under which the society would exhibit mixed opinions and for 
conditions of final uniformity or near uniformity (either almost all $+$ or almost all $-$).
In particular, we were looking for conditions supporting highly asymmetric configurations, for example
$N_+=420$, when a small number of $-$ fanatics could be `surrounded' by some opportunist followers -- situation corresponding to many important social issues, such as support for extremist views.

The main control parameters were the initial number of $+$ opinion holders, $N_+^{ini}$, the external influence $h$ and the ratio between the average number of static 
($S$) and free links ($F$)  per agent. In addition to $N_+$, we have studied the rate of convergence,
the distribution of opinions for the opportunistic and fanatic agents, and 
averages of the numbers of links between various types of agents.

\begin{figure*}
\includegraphics[scale=0.65]{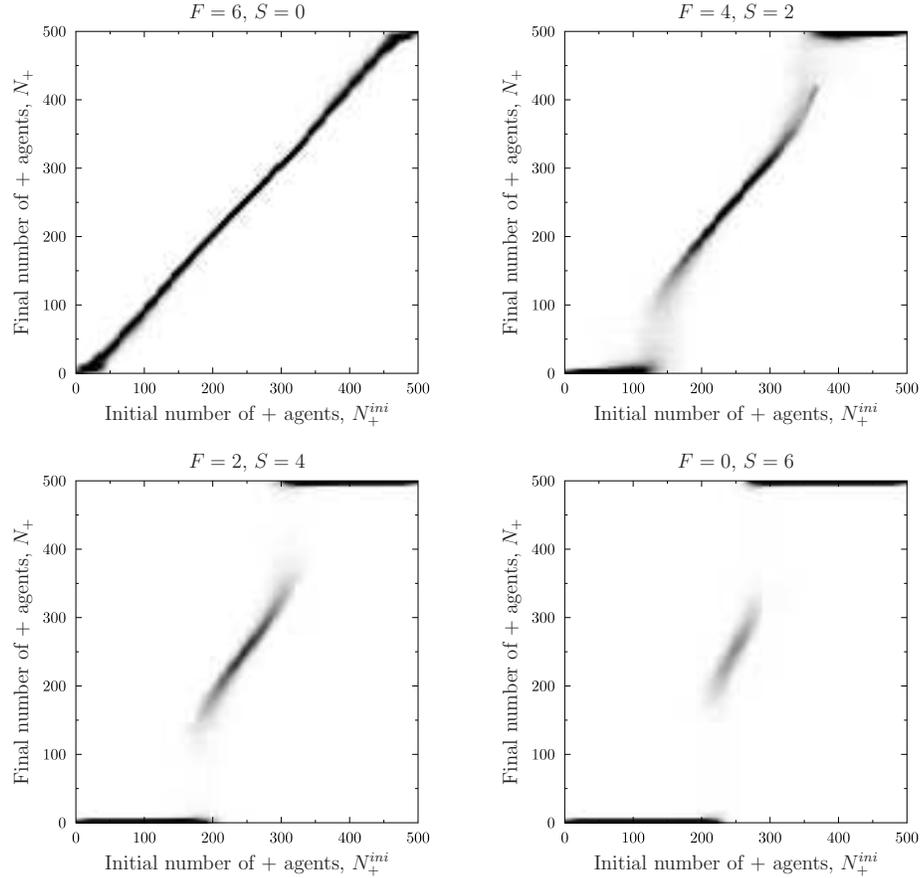} 
\caption{Probability distributions of final $N_+$ values  as functions of their initial number $N_+^{ini}$ for external influence $h=0.0$ .  \label{H0000}  
The panels correspond to various mix of average free and static links per agent.
The simulations started with the number of fanatic agents making on average 10\% of the given starting value of $+$ and $-$ agents. The fanatics had medium switching probability of $\omega_f = 0.02$ and $d=0.5$ (meaning that roughly 50\% of the social influence was due to nearest neighbors).
}
\end{figure*}

Accumulation of results of thousands of individual simulations has revealed some surprising patterns in the system behavior. Figure~\ref{H0000} presents probability distributions of simulated societies for various combinations of free and static links and for external influence $h=0$, $\omega_f=0.02, \omega_o=0.9$ and $d=0.5$.  Gray density corresponds to relative probability of finding a simulation ending with a particular value of $N_+$ for a simulation starting with a given $N_+^{ini}$ value. Let us start with the `traditional' model, in which the network structure is frozen (lower right panel, $F=0, S=6$). As we can see, the mixed society prevails only for a very limited range of initial values of $N_+^{ini}$, between 230 and 270. Any larger asymmetry in the initial distribution leads to societies with uniform distribution, favoring the initial majority. In the central region, the final distribution has a strong variance, but the peak of the distribution lies along a line with a slope of $5/3$. The transition between the mixed and uniform solutions is rather abrupt, suggesting a phase transition between the random, mixed state and a full consensus.

When we turn to the opposite end of the network flexibility, namely the simulations where every link is free to be cut (upper left panel, $F=6, S=0$), we see a  totally different behavior. First observation is that the results of the simulations show much less variance between runs. But more important is the fact that for almost the whole range of starting conditions the society composition remains almost unchanged during the `opinion switching', $N_+ \approx N_+^{ini}$. Thanks to separation of the opposing agents during the preparatory phase, even the influence of a large majority is weakened by social distance introduced by the severed links. 

The two intermediate cases ($F=2, S=4$ and $F=4, S=2$) are similar to the static society. The points of transition to uniform opinion distribution shift to more extreme values of $N_+^{ini}$ with increasing fraction of free links, but still the absence of final configurations with a large majority/small minority is clear.

\begin{figure*}
\includegraphics[scale=0.65]{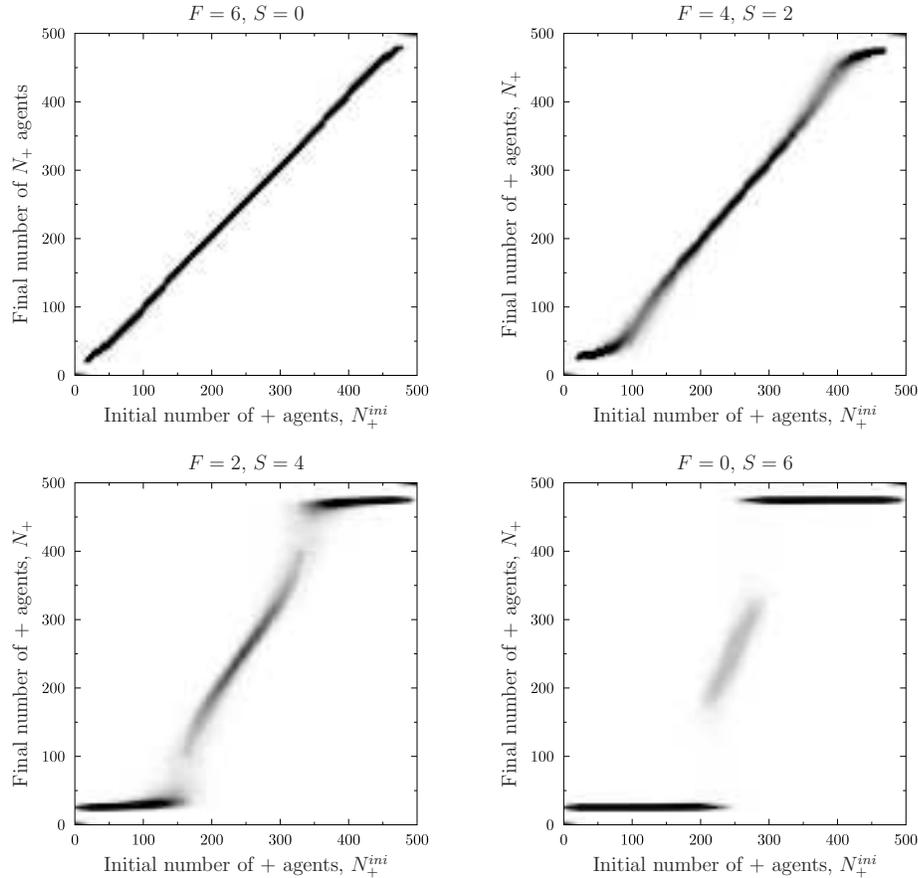} 
\caption{Probability distributions of final $N_+$ values  as functions of their initial number $N_+^{ini}$ for external influence $h=0.0$ .  \label{9H0000}  
The panels correspond to various mix of average free and static links per agent.
The simulations started with fixed number of 25 fanatic $+$ agents, with very low switching probability of $\omega_f = 0.000001$ and $d=0.3$ (meaning that roughly 70\% of the social influence was due to nearest neighbors).
}
\end{figure*}

To show how this behavior is influenced by the simulation parameters we present results obtained for a smaller value of $d=0.3$. Such choice leads to much higher influence of immediate neighbors. Moreover, we changed the $\omega_f$ value to $0.000001$, making it virtually impossible to switch the opinion of a fanatic. We have also changed the procedure for initial assignment of opinions of fanatics. In the previously presented simulations the initial number of fanatics with a given opinion was taken as 10\% of the general initial assignement (i.e. for $N_+^{ini}=100$ there were, on average 10 $+$ fanatics). Here, regardless of the $N_+^{ini}$, we kept a constant number of 25 $+$ fanatics and the same for their $-$ counterparts. The aim of this change was to study the effects that a sizeable number of fanatics may have on a society, even if their opinion are not supported by opportunistic

The overall changes brought by these adjustments were quite interesting (Fig.~\ref{9H0000}). The two limiting (all-free and all-static) cases retained their characteristics. The linear behavior of the $F=6$ case is well preserved, as is the rapid transition in the $F=0$ case. The change in the values of $N_+$ in the `consensus' regime was due to minority fanatics who, by design, did not change their opinions, but who did not have any supporters. 

The most interesting was the intermediate configuration with $F=4$. Increasing the role of the nearest neighbors has led to change from discontinuous behavior to a continuous one. The stronger influence of immediate neighborhood and capability to adjust, on average, 66\% of the links, resulted in societies where in addition to fanatics, some other agents could keep their minority opinions.

\begin{figure*}
\includegraphics[scale=0.65]{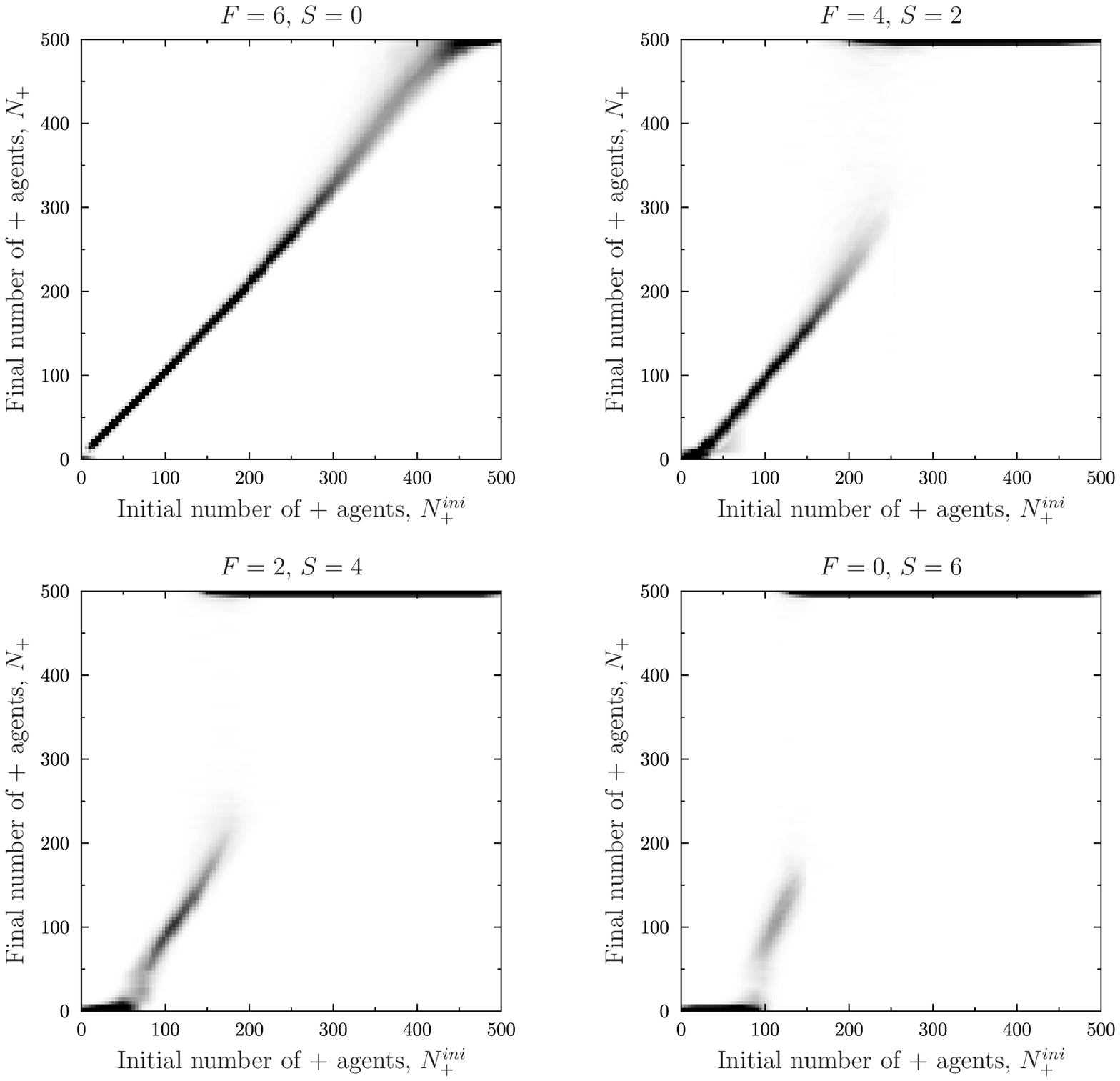} 
\caption{Probability distributions of final $N_+$ values  as functions of their initial number $N_+^{ini}$ for external influence $h=0.5$ .  \label{H0500}  
The panels correspond to various mix of average free and static links per agent.
The simulations used the same parameters ad Fig.~\ref{H0000}.
}
\end{figure*}

The next set of results (Fig.~\ref{H0500}) presents simulations where external influence was present ($h=0.5$). The rest of the simulation parameters was identical to those presented in Fig.~\ref{H0000}.
For the $F=6$ case, the presence of positive influence has stabilized the initial distribution for small values of $N_+^{ini}$, while at the same time broadening the distribution of possible outcomes for $N_+^{ini} > 300$, where simulation outcomes between $N_+^{ini}$ and 500 were possible. It should be noted, however, that in this case is was still possible to find final societies starting with a minority of a $100$ $-$ agents and ending with a similar value, despite being outnumbered and under external influence.

For the cases where at least some of the links were static, the  general opinion followed the external influence  for a broad range of starting conditions. For networks with mixed free and static links the ranges of nonuniform opinion results were shifted to smaller starting values of $N_+^{ini}$, and the transitions to consensus were abrupt. In the case of a fully static network, the external influence was too weak to break negative consensus for starting conditions where there were more than 400 agents with negative opinions at the starting point. Only for a very narrow range of conditions we observed mixed opinion state, and above $N_+^{ini}=140$ a positive consensus prevailed.

\begin{figure*}
\includegraphics[scale=0.65]{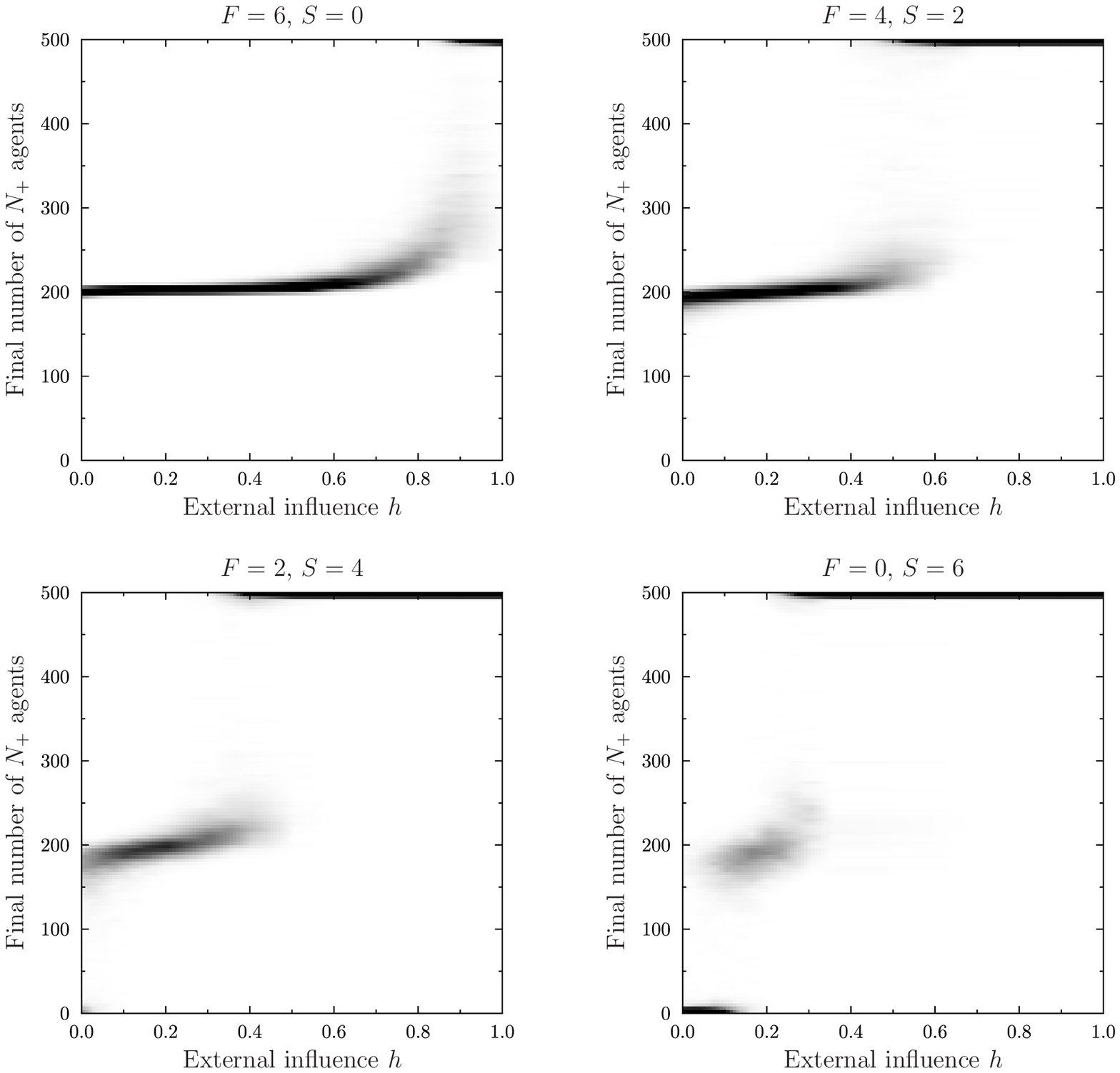} 
\caption{Probability distributions of final $N_+$ values  as functions of external influence for various ratios of $F$ and $S$, and for starting value of $N_+^{ini}=200$. Simulation parameters are the same as in Fig.~\ref{H0000}. \label{NH}
}
\end{figure*}

To test the importance of the external influences we have run a series of simulations keeping the initial number of $+$ agents constant (using slightly asymmetric value of $N_+^{ini}=200$), but with different values of $h$, between $0$ and $1$. Results are presented in Fig.~\ref{NH}. 
For fully flexible network, the final composition was remarkably stable against the external influence. Its effects became visible above $h=0.6$ and around $h=0.8$ led to swift transition to consensus state.
With increasing ratio of fixed links the region of stability has been reduced, but still the average final composition of the society remained almost unchanged for a broad range of external propaganda.
This is in contract with the fully static, traditional network model, where the range of values leading to mixed society was very narrow.

\section{Comparison with other models of consensus on dynamic networks}

Recently there is increasing interest on opinion modeling on dynamic social structures, as opposed to more traditional static ones. There are numerous examples of related works, using different social networks and, importantly, different models for opinion change of individual agents.  
Examples of such works are provided by 
Holme and Newman\cite{16},
Kozma, Barrat and Nardini\cite{17,18}, 
Vazquez, Eguíluz and San Miguel\cite{19} and  
Fu and Wang\cite{20}.
In these works, in each iteration, an agent with certain probability $\phi$ would adjust its opinion (in accordance with the chosen model) and with probability $1-\phi$ would cut one of the links with a neighbor of a different opinion. All the cited works start with random networks. The differences in approaches lie partially in the rewiring process and partially in the opinion model.
In [\citep{16,17,18}] and [\citep{19}] the new links are formed to random agents, while in \citep{20} there is preference for the next nearest neighbors ($S_2$ in our notation). Much greater differences are in the opinion descriptions. In [\citep{16,20}] there are multiple opinion values, and the final state of the system is a set of separated opinion islands, each corresponding to a given value of opinion. While  Holme and Newman\cite{16} use very simplistic rule, in which the initiating agent simply takes the opinion of the target agent, Fu and Wang\cite{20} use majority rule among nearest neighbors. The remaining works model opinions via the Voter model (\citep{18,19}) or Deffuant bounded confidence model (\citep{17}).

Compared to these works the main new factors bought by our approach are in the opinion dynamics. While the effects of network changes, leading to separation of agents holding differing opinions are generally similar, the inclusion of weighted influence by the whole society marks a significant change from the focus on individual convergence of opinions.
We admit, that the specific form of splitting the society and assignment of weights to agents in various groups through social distance is arbitrary and more complex than in the individual encounter models. 
Yet there is, in our opinion, a strong motivation for making this effort, based on observations of rarity of situations when people change their opinions due to individual encounters. In fact, there are stable social networks based on purely negative connections, where repeated contacts lead to entrenchment in the opposing opinions without any chance of lessening differences of opinions\cite{21}.
We hope that the control parameters, even though more numerous,  have more direct correspondence to factors measurable by sociologists.

Obviously the first important factor is the ratio of free to static links. The works we referred to in this section use assumption that all the links can be rewired, which is obviously in disagreement with many social situations. It is worth noting that recently Lambiotte and Ausloos\cite{22} have proposed a majority rule based model on static coupled random networks. Such networks bear close resemblance to the final state of our rewiring process, and the authors have studied dependence of the outcomes on the ratio of links connecting the coupled random subnetworks. Despite different opinion switching model, the results are close to those obtained by us, namely, it is possible to retain mixed opinion state only when the number of links between the subnetworks is sufficiently small (in our language, if $S$ is small).

\begin{figure*}
\includegraphics[scale=0.65]{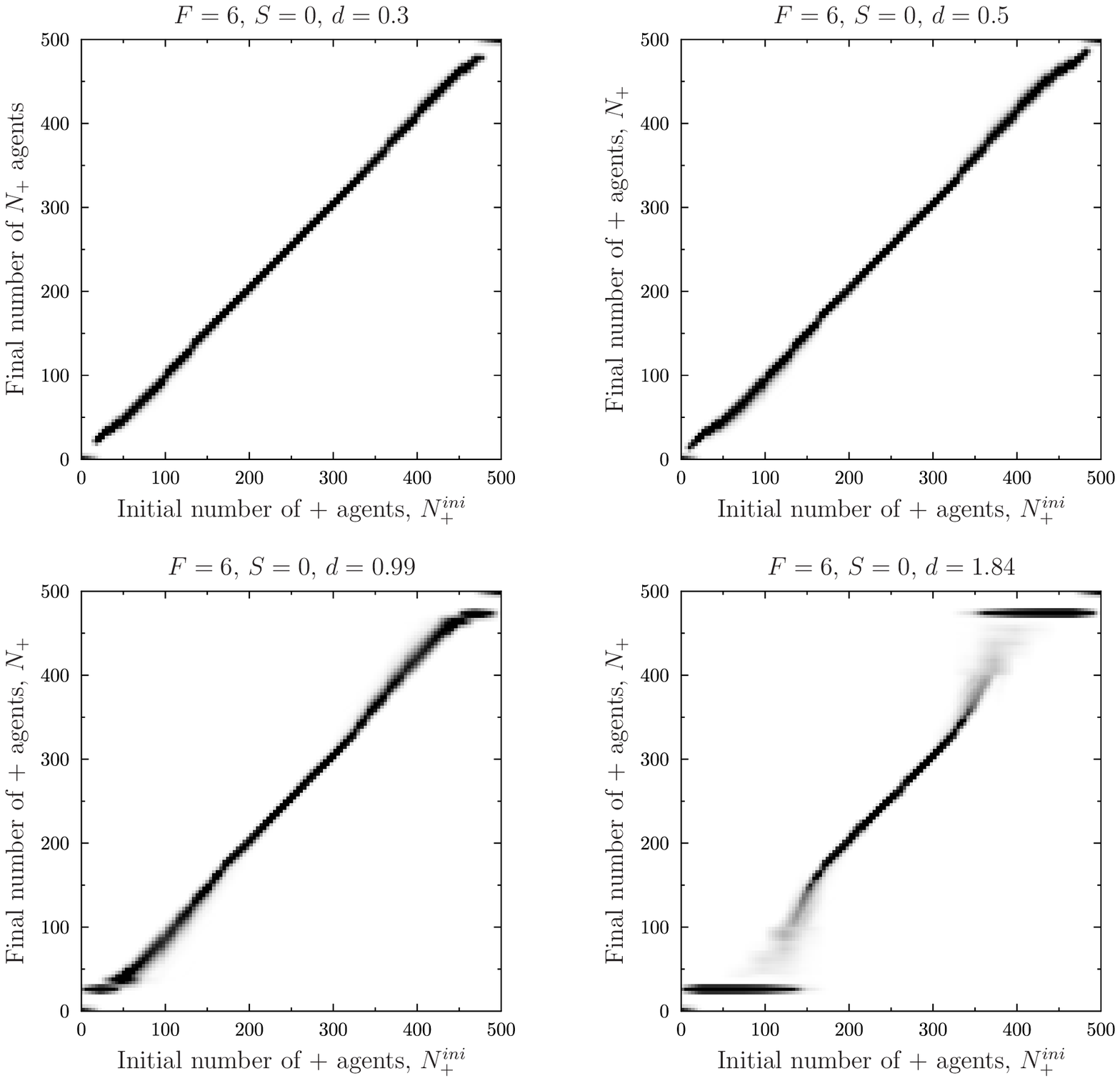} 
\caption{Probability distributions of final $N_+$ values  as functions of parameter $d$ for networks with only free links.  The simulations started with fixed number of 25 fanatic $+$ agents, with very low switching probability of $\omega_f = 0.000001$. Low values of $d$ correspond to high importance of nearest neighbors, while for $d=0.99$ about 25\% of the influence comes from the each of the $S_K$ spheres. For $d=1.84$ the opposite effect is in force: 50\% of the influence comes from $S_4$, i.e. furthest strangers.   \label{DX}
}
\end{figure*}

Another example of the value of the extended model is in the role of parameter $d$, which determines the relative role of the closest neighborhood on social influence. As we observed, decreasing $d$ (which means that the role of the nearest neighbors is higher) brings more stability to asymmetric configurations with large majority and small minority. 
On the other hand, when the value of $d$ is increased, for example to 0.99 (when the influence of each $S_K$ sphere is roughly identical) or to $d=1.84$, when half of the influence originated from the most remote parts of the society, the system behavior starts to resemble closely situation with significant number of fixed links (see Fig.~\ref{DX}). This is understandable because $S_4$ influence is effective whether there is separation between the two subnetworks or not. Thus the parameters' role is, to certain degree, interchangeable.

The last of the elements influencing the whole system is the presence and number of fanatics. In a recent preprint, Galam\cite{23} has used similar concept of `inflexible agents'. It is interesting, however, that within his model, which does not allow social network modification, even inclusion of small number of fanatics against huge majority favoring the other opinion would eventually lead to the whole population aligned with the fanatics. As we have shown, the coupling of the presence of fanatics and the active status of the network leads, at best, to the continued existence of minorities, but never to the `dictatorial machine' of dominating minority.

\section{Comparison with real world phenomena}

\begin{figure*}
\includegraphics[scale=0.7]{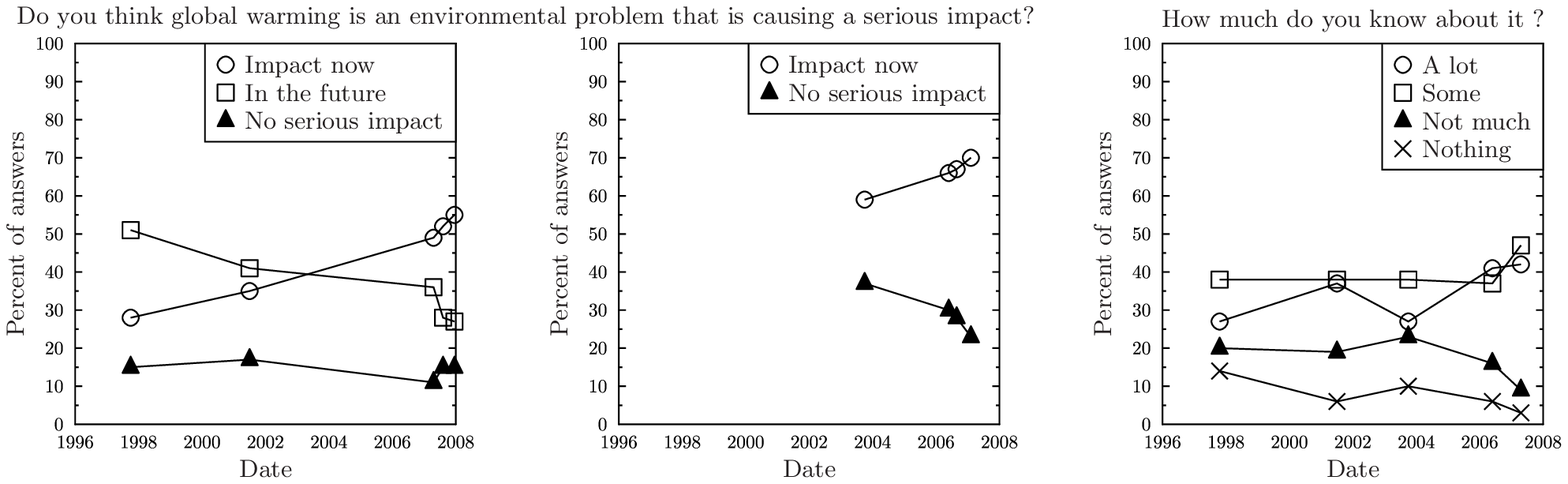} 
\caption{Evolution of opinions on global warming in US, according to CBS polls. The stability of the minority view, especially in the light of massive public propaganda is remarkable. \label{CBS-GW}
}
\end{figure*}

For comparison with real world data we have chosen opinions on the severity of threat posed by global warming (Fig.~\ref{CBS-GW}). The presented data are based on CBS polls (\url{http://www.cbsnews.com/stories/2007/10/12/politics/main3362530.shtml}) and Gallup data\cite{24}. The relative stability of the number of `global warming deniers' is especially interesting in the light of the massive amounts of propaganda on the subject. The growth of awareness (shown in the third panel of Fig.~\ref{CBS-GW}) does not seem to lead to a consensus. One of the reasons might be the chasm separating the two groups. This is visible, for example in the  studies of the reactions to the environmental propaganda\cite{24}.  While on average 41\% of Gallup respondents think that the seriousness of the global warming is exaggerrated (percentage which grew from 30\% in 2006), the ratio is only 22\% for Democrat supporters and 66\% for Republicans. 
As we have seen,  the links between these two camps are largely broken, so the situation corresponds at least partially to our model. Even in scientific analyses of global warming, a lot of discussions related to importance of human driven factors in the climate change are often shifted to emotional grounds and arguments \emph{ad-personam}. As a result, the gap between the communities only broadens, with each of the groups following their own gurus and leaders, without any effects of gradual consensus building. We expect that the `deniers' would be present in societies even when overwhelming, direct evidence will accumulate, clustering among any evidence pointing to natural origins of the climate changes and distrusting all other information sources.

It is interesting that Galam\cite{23} uses global warming as a test case for his model (which, as already noted, does include fanatics but no dynamic network changes). By putting scientists (or people who claim that there is scientific proof for human culpability for the phenomenon) in the role of the fanatics, he predicts that there are conditions under which we would achieve a total victory of this opinion. Thus, the two models predict significantly different outcomes. The observed stability of the global warming deniers shows, in our opinion, the importance of the voluntary separation on general opinion statistics.

Another example of a stable, highly polarized opinion split is provided by pro-choice and pro-life approaches to abortions. The percentages of those supporting and rejecting abortion rights in the US remains almost unchanged for the past 12 years, as documented by the CBS polls.  Especially interesting are data from two polls from 1998. The first indicated that 44\% of respondents thought that people advocating making all abortions illegal are unreasonable extremists. This would make the expectation for achieving consensus minimal. Indeed, in another poll,  88\% of respondents did not believe that the disagreement shall ever be resolved.  While these data are based on US statistics, the Polish society is equally split both in opinions and in the ability to form any communication channel between the two camps, which simply throw abuse at the other's positions. Achieving opinion consensus in such case is, indeed, quite impossible. 

Important examples of the stability of mixed opinion  may be provided by the situations of intense social conflict (such as Afghanistan or Northern Ireland) where despite the efforts for reconciliation, the existence of small number of fanatics and cut-off of links between the proponents of conflicting views results in almost unbreakable deadlock. The simple model used in our simulations shows how probable are societies where a small number of fanatics can attract enough supporters to provide self-sustained sub-communities, separated from the rest of society. It would be very interesting to see studies of real social links in these communities, especially for the periods when the violence was significantly reduced, as was the case of Northern Ireland.

An interesting analysis of opinion formation showing persistence of opinions despite massive propaganda effort has been presented by Wragg\cite{25}. His analysis of support for polio vaccination in Uttar Pradesh rural communities combines social distance (in Nowak-Latan{\'e} spatial model) and factors related to group separation due to religion. Key factor in refusal of vaccination has been attributed to rumors among the Muslim part of the population    that the vaccinations were designed to sterilize Muslim children as part of a  well-disguised  form  of  genocide  implemented  by  the  Hindu-dominated  Indian  Health Department. The split between religious groups has affected the acceptance  this rumor, and subsequently, the reactions to vaccination campaigns.

Due to limitations of compute power, our simulations were limited to small networks (500 agents). However, even such size might be relevant and compared with properly chosen social observations. Examples of small, relatively isolated communities are numerous. For example, one could consider reactions of company workers to political issues (such as the classical study of Berelson,  Lazarsfeld, and  McPhee\cite{26}, where analysis of voter co-workers and friends shows significant preference for same-opinion links\footnote{Chart XL of Ref.~\citep{26} shows that among republican voters 55\% had all three closest fiends being republican (47\% ratio for the democrats) and as much as 45\% in the case of co-workers, where the control over one's associations is smaller (27\% in the case of democrat voters). The voters strength of convictions has been found related to the homogeneity of his associates. Moreover, the people were found to discuss politics mostly with people like themselves. The authors provide quotes such as `All my neighbors are Democrats, so there's no use talking with them about it'. From a personal perspective I can only confirm that this situation remains true fifty years later in a country halfway across the globe from US.}) or, even more interestingly to vital company issues, such as privatisation of state enterprises.  Or, looking for more flexibility in link formation and destruction, to study internet social networks. More traditional local communities, especially in rural areas, offer similar scale and the issues in question might involve adoption of innovations  or political or ethical issues -- subject of extensive study in sociology.

The persistence of minority groups, separated from the rest of society is observed on many occasions. It plays especially important role in areas of social conflict, where the consequences of opinion split are often disastrous, leading to terrorism. The separation between minority opinion holders and the majority results in inability to convince them and to separate the more flexible members of the minority from the fanatics. For these reasons we believe that the current model, stressing the dynamic nature of social bonds and interactions can reproduce opinion formation in many social situations. 
While the fact that there are relatively many control parameters might seem as a disadvantage of the model, we believe that they can be derived from direct social research for specific communities and situations. Such experimental determination is possible for group size, average number of links, ratio of free and static links or the degree distribution. Also the tendency to act as the `fanatics'  is measurable via interviews, giving the estimates for $o_o$ and $o_f$. Similarly, the parameter $d$ could be estimated from the way people assign importance to opinions of their close acquaintances and to more remote parts of society. As a result, the additional complexity of the model might be less daunting that it initially looks, and allow specific predictions for concrete social situations, something that is difficult in the simpler approaches.
We would be most interested in social studies corresponding to the modeled situations, as the proof of the model lies only in its ability to reflect reality.
Future directions of our research include simulations with much larger societies and simulations where a  society composition resulting from a simulation is \emph{destabilized}  by significant change of external influences.


\end{document}